\documentclass[]{fairmeta}
\usepackage{amssymb}
\usepackage{txfonts}
\usepackage{tabularx}
\usepackage{booktabs} 

\usepackage{makecell}

\usepackage{xspace}

\allowdisplaybreaks[4]

\usepackage{amsmath,amsthm,amscd,dsfont,mathrsfs,mathtools,microtype,nicefrac,pifont,soul}
\usepackage[utf8]{inputenc}
\usepackage[T1]{fontenc}   
\usepackage{url}           
\usepackage{booktabs}     
\usepackage{algorithm,algpseudocode}
\usepackage{bbm}
\usepackage{bm}
\usepackage{color, colortbl}
\definecolor{LightCyan}{rgb}{0.88,1,1}
\usepackage{dirtytalk}
\usepackage{enumerate}
\usepackage{graphicx}
\usepackage{listings}
\usepackage{subfigure}
\usepackage{enumitem}
\usepackage{xspace}
\usepackage{makecell}
\usepackage{multirow, array}
\usepackage{verbatim}
\usepackage{tikz}
\usepackage{tcolorbox}
\usepackage{tablefootnote}
\usepackage{cancel}
\usepackage[symbol]{footmisc}
\renewcommand{\thefootnote}{\fnsymbol{footnote}}

\usepackage{caption}
\usepackage{wrapfig}
\newtheorem{theorem}{Theorem}

\newtheorem{definition}{Definition}

\newcommand{\cB}{\mathcal{B}}

\newcommand{\cD}{\mathcal{D}}

\newcommand{\cO}{\mathcal{O}}

\newcommand{\T}{^\top}

\newcommand{\E}{\mathbb{E}}

\newcommand{\norm}[1]{\left\|#1\right\|}

\DeclareMathOperator*{\argmin}{argmin}

\def\<#1,#2>{\left\langle #1,#2\right\rangle}
\mathchardef\mhyphen="2D
\allowdisplaybreaks

\newcommand{\R}{\operatorname{\mathbb R}}

\newcommand{\range}{\operatorname{range}}

\DeclarePairedDelimiterX{\infdivx}[2]{(}{)}{%
  #1\;\delimsize\|\;#2%
}

\theoremstyle{plain}          
\newtheorem{thm}{Theorem}[section]
\newtheorem{lma}[thm]{Lemma}

\theoremstyle{definition}

\title{MultiBalance: Multi-Objective Gradient Balancing in Industrial-Scale Multi-Task Recommendation System}
\author{Yun He$^*$, Xuxing Chen$^*$, Jiayi Xu, Renqin Cai, Yiling You, Jennifer Cao, Minhui Huang, Liu Yang, Yiqun Liu, Xiaoyi Liu, Rong Jin, Sem Park, Bo Long, Xue Feng}

\abstract{In industrial recommendation systems, multi-task learning (learning multiple tasks simultaneously on a single model) is a predominant approach to save training/serving resources and improve recommendation performance via knowledge transfer between the joint learning tasks. However, multi-task learning often suffers from negative transfer: one or several tasks are less optimized than training them separately. To carefully balance the optimization, we propose a gradient balancing approach called MultiBalance, which is suitable for industrial-scale multi-task recommendation systems. It balances the per-task gradients to alleviate the negative transfer, while saving the huge cost for grid search or manual explorations for appropriate task weights. Moreover, compared with prior work that normally balance the per-task gradients of shared parameters, MultiBalance is more efficient since only requiring to access per-task gradients with respect to the shared feature representations. We conduct experiments on Meta's large-scale ads and feeds multi-task recommendation system, and observe that MultiBalance achieves significant gains (e.g., 0.738\% improvement for normalized entropy (NE)) with neutral training cost in Queries Per Second (QPS), which is significantly more efficient than prior methods that balance per-task gradients of shared parameters with 70\raisebox{-0.9ex}{\~{}}80\% QPS degradation.}

\date{\today}
\correspondence{Yun He (\href{mailto:yunhe2019@meta.com}{yunhe2019@meta.com})}

\begin{document}

\affiliation{Meta}
\maketitle
\def\thefootnote{*}\footnotetext{Equal contributions.}\def\thefootnote{\arabic{footnote}}

\section{Introduction}
Multi-task learning (MTL) \citep{ruder2017overview, vandenhende2021multi} is to optimize multiple task objective functions simultaneously on a single model. In a typical MTL architecture, there is a shared bottom layer to learn a feature representation via mapping categorical features and numerical features to embeddings and then conducting feature interactions. Each task has a head (e.g., a multi-layer perceptron) which receives the feature representation from the bottom layer and learn to minimize the task loss function. Multi-task learning has been applied in a variety of areas such as computer vision \citep{kendall2018multi}, language understanding \citep{liu2019multi} and recommendation systems \citep{wang2018explainable}.

In particular, multi-task learning is important for many industrial recommendation systems due to two reasons: (1) Saving serving resources. There are billions of users in platforms like Amazon, Meta and YouTube. Multi-task learning enables one single model to serve multiple tasks for these users, saving huge amounts of machines and electricity consumption; (2) Knowledge transfer. Normally, semantically related tasks are co-trained together where knowledge is transferred among each other to obtain a better performance.

The most straightforward way to optimize a multi-task model is to minimize the sum of per-task losses. Although knowledge transfer is ideally helpful, multi-task learning often suffers from negative transfer \citep{kanakis2020reparameterizing, ruder2017overview, zhang2022survey}, which refers to the worse performance on a task than learning it separately, caused by the co-learning of other tasks. A possible root cause is gradient conflicts such as one task gradient with respect to (w.r.t.) the shared parameters are so large that they dominates the optimization. Therefore, we need to carefully balance the joint learning of multiple tasks.

In industry, a classic way to balance multi-task learning is to multiply each task loss with a task-specific weight and then minimize the weighted-sum loss. Being simple and effective, however, it could be quite time-consuming to search for an appropriate set of weights, especially when the task weights are obtained via grid search, random search or manual search, and the search space grows exponentially with the number of tasks. Moreover, across the upgrades of a model, the setting also changes: a new task might be introduced or a old task might be deprecated, then the previous weights might no longer be optimal (and could be even harmful sometimes) and the weights need to be re-searched. Therefore, a research question arises here: how to automatically learn the task weights during the model training to balance the optimization?

In recent years, casting multi-task learning as a multi-objective optimization (MOO) problem emerges as an important method to achieve Pareto optimal solution. Among provably convergent ones, a classic method is multi-gradient descent algorithm, \texttt{MGDA} \citep{sener2018multi}, which learns the task weights via minimizing the norm of convex combination of per-task deterministic gradients. Recently, the variants of \texttt{MGDA} for stochastic gradients like \texttt{MoCo} \citep{fernando2022mitigating} are proposed. However, these MOO experiments are originally conducted on small models but less explored in real industrial large-scale systems with billions of model parameters, where both performance and efficiency are highly important.

There are two challenges for applying MOO in industrial systems: (1) MOO usually requires access to per-task gradients w.r.t. the shared parameters, which will pose significant computation and memory cost on large-scale systems; (2) Variance in stochastic gradients will incur potentially large instability in update direction of model parameters~\citep{liu2021stochastic, fernando2022mitigating, xiao2023direction}, and this issue becomes more severe when it comes to industrial-scale systems. Hence, how to stabilize the training of models when MOO meets stochastic gradients in industrial systems is important and challenging.


To overcome these challenges, we propose \texttt{MultiBalance} as a practical solution to balance gradients in large-scale multi-task recommendation systems. We have the following contributions:
\begin{itemize}
    \item To efficiently balance gradients in large-scale systems, MultiBalance balances per-task gradients w.r.t. the shared feature representation (the output of the shared bottom layer) rather than balancing per-task gradients w.r.t. the shared parameters as in most of previous work. We are aware of the fact that prior works \citep{sener2018multi,liu2021towards,javaloy2021rotograd} have demonstrated that balancing gradients of the shared feature representation is efficient but their experiments were conducted on small models and small public datasets. To our best knowledge, this paper is the first one reporting the successful adoption of MOO in industrial-scale systems.
    \item To stabilize the training of large models, MultiBalance maintains a moving average of the magnitude of the mini-batch stochastic gradients throughout the training process. 
    \item We analyze the theoretical characteristics of feature representation gradients and show it is a good surrogate of parameter gradients. We also give a condition when MultiBalance is provably to achieve Pareto stationary point.
    \item We significantly increase the efficiency of gradient balancing and successfully apply it on Meta’s large scale multi-task recommendation system, which proves to be a high ROI solution for large scale industrial systems. In particular, we observe that MultiBalance achieves significant gains (e.g., 0.738\% improvement for normalized entropy (NE)) with neutral training cost in QPS, which is significantly more efficient than prior methods that balance per-task gradients of shared parameters with 70\raisebox{-0.9ex}{\~{}}80\% QPS degredation.
\end{itemize}

\section{Related Work}
\smallskip
\noindent\textbf{Multi-task Learning in Recommendations.} Multi-task learning has been widely applied to model various user behaviors in recommender systems. \citet{lu2018like} and \citet{wang2018explainable} design multi-task models to jointly optimize the recommendation task and the corresponding explanation task. \citet{hadash2018rank} propose a multi-task framework to learn the items ranking task and rating task together. Multi-task learning is also used in video recommendations \citep{zhao2019recommending, tang2020progressive} to optimize different objectives simultaneously. 

\noindent\textbf{Loss Balancing.} One important technique in MTL is to balance the weights of different tasks based on their objective function values. In particular, the task weights are set to be trainable during the whole training process and adjusted according to pre-defined criteria. Notable works include uncertainty weighting~\citep{kendall2018multi}, learnable loss weights~\citep{liebel2018auxiliary}, and impartial learning~\citep{liu2021towards}.

\noindent\textbf{Gradient Balancing.} Different from loss balancing methods that may not handle gradient conflicts, another line of work proposes gradient balancing to address this issue. For example, by utilizing the information obtained from task-specific gradients, one can conduct gradients re-weighting~\citep{desideri2012multiple, fliege2019complexity, liu2021conflict, liu2021stochastic, fernando2022mitigating}, gradients manipulation such as projection (\texttt{PCGrad}~\citep{yu2020gradient}), and adjusting the gradient direction (\texttt{Gradient Vaccine}~\citep{wang2021gradient}). Among these methods, a widely-used method with strong convergence guarantees is multi-gradient descent algorithm (\texttt{MGDA}), which carefully designs the update direction of model parameters to achieve Pareto stationarity. Deterministic version of multi-gradient descent algorithm has been studied in \citet{desideri2012multiple} and \citet{fliege2019complexity}. In each iteration, \texttt{MGDA} searches for the convex combination of gradients that has minimum norm as the update direction, and then update model parameters along according to it. \texttt{MGDA} is provably capable of finding a Pareto stationary point \citep{fliege2019complexity}. Some follow-up works are dedicated to more general settings such as handling bias caused by stochastic gradients~\citep{liu2021stochastic, fernando2022mitigating, xiao2023direction}, generalization bound~\citep{chen2023three}, etc. It is worth noting that, computing all the task-specific gradients of model parameters requires multiple backward passes that greatly slow down the training. 

\noindent\textbf{Key Difference.} Although a few prior works consider applying balancing algorithms to representation-level gradients~\citep{sener2018multi,liu2021towards,javaloy2021rotograd}, their experiments are conducted on small models and small public datasets. Besides, there exist cases in which balancing algorithms applied to representation-level gradients do not perform well~\citep{navon2022multi}. Thus the applicability of such methods in industrial-scale systems remains unknown. To our best knowledge, this paper is the first one reporting the successful adoption of MOO-based gradient balancing in industrial-scale systems.

\section{Preliminaries}
\subsection{Industrial Multi-Task Ranking System}\label{sec: MTL_industry}

\begin{figure}[h]
  \centering
  \setlength{\abovecaptionskip}{0.0cm}
  \setlength{\belowcaptionskip}{0.0cm}
  \includegraphics[width=0.7\linewidth]{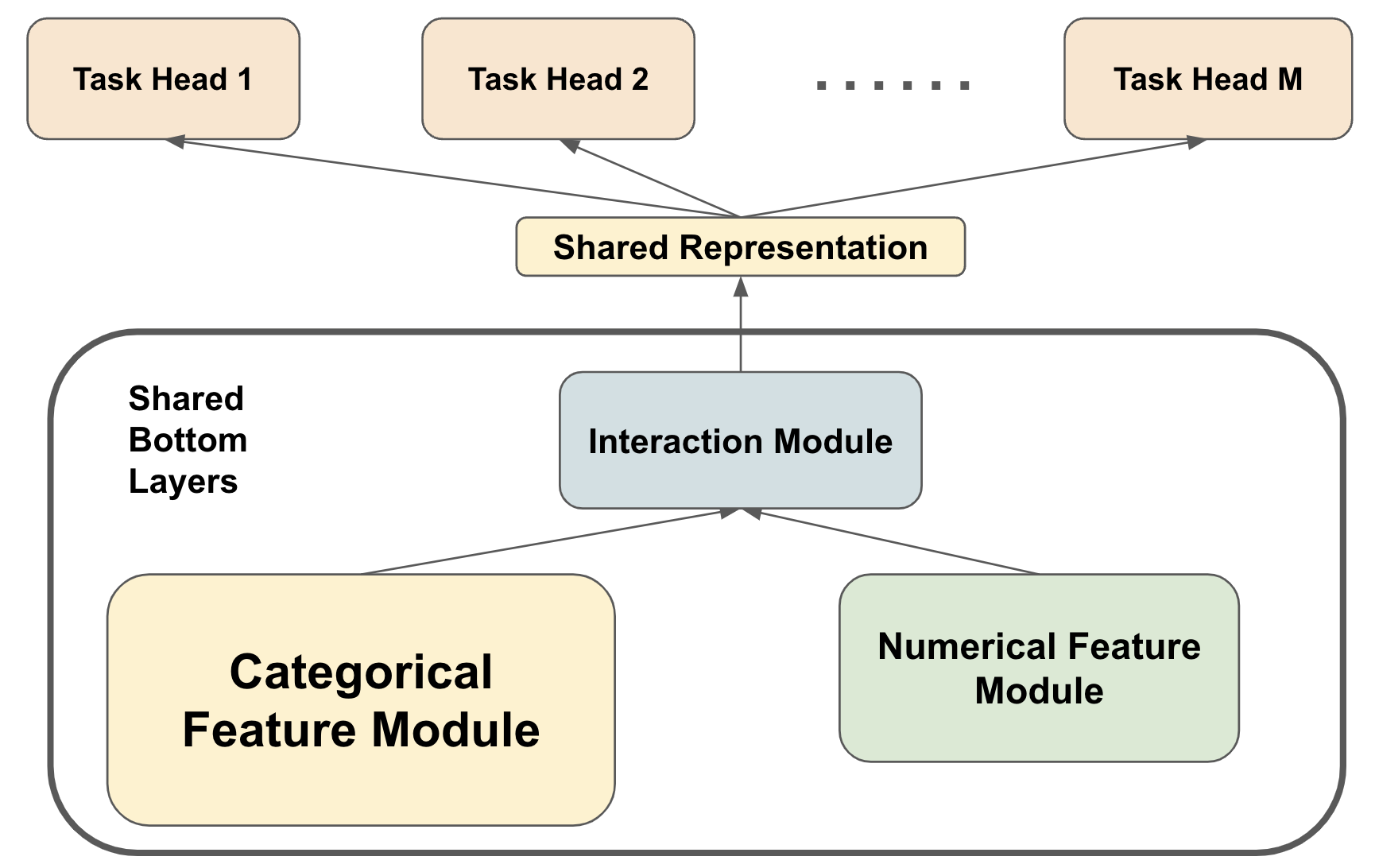}
  \caption{General Framework of Multi-Task Ranking Model.}
\label{fig: SparseNN}
\end{figure}

Figure \ref{fig: SparseNN} illustrates the general framework of an industrial multi-task ranking model \citep{wang2023towards, li2023adatt}, which includes three parts: shared bottom layers, a shared representation and task heads.

\textbf{Shared Bottom Layers.} This part learns the feature representation from the model input -- data samples' categorical features and numerical features. On one hand, there is a categorical feature module using embedding table lookups to map one-hot/multi-hot sparse features into embedding vectors. On the other hand, a numerical feature module maps numerical features into embedding vectors. After that, an interaction module (e.g., DeepFM \citep{guo2017deepfm}, DCN \citep{wang2017deep}) models the interaction between features. 

\textbf{Shared Representation.} The output of the shared bottom layers is a two-dimensional feature representation tensor, where first dimension is the batch size and the second dimension is the same as dimension of the last layer of the shared bottom layers. 

\textbf{Task-specific Heads.} At last, the shared feature representation is fed into a task module (e.g., a multi-layer perceptron) for each task to calculate the objective loss. 

\textbf{Categorical Feature Module Contributes Most Parameters.} A characteristic of industrial ranking model is that most parameters come from categorical feature module. This is understandable in the industrial world: each categorical feature might have millions or billions of possible categories (e.g., item id). Therefore, there is a large quantity of high-dimensional embedding tables. \citet{jiang2019xdl} release the parameter statistics of three public models used in their daily work. As shown in Table \ref{tab:Proportion of parameters}, we observe that categorical feature module has more than 90\% (or even more) parameters.

\begin{table}[htbp]
  \centering
      \setlength{\abovecaptionskip}{0.0cm}
  \setlength{\belowcaptionskip}{0.0cm}
  \small
  \caption{Proportion of Parameters within Categorical Feature Module at Three Public Models}
    \begin{tabular}{lccc}
    \toprule
    Models & DNN \citep{huang2013learning}   & DIN \citep{zhou2018deep}  & Crossmedia \citep{ge2018image} \\
    \midrule
    \#Param of Other Modules & 1.2 Billion & 1.7 Billion & 1 Million \\
    \#Param of Categorical Module & 18 Billion & 18 Billion & 5 Billion \\
    Ratio of Categorical Module  & 93.75\% & 91.37\% & 99.98\% \\
    \bottomrule
    \end{tabular}%
  \label{tab:Proportion of parameters}%
\end{table}%



\subsection{Multi-Objective Optimization for Multiple Co-Trained Ranking Tasks}
\label{sec: preliminary of Multi-Objective Optimization}
From an optimization perspective~\citep{sener2018multi}, we may view the multi-task learning in Section \ref{sec: MTL_industry} as a multi-objective problem. Mathematically, we consider the following multi-objective optimization problem.
\begin{align}\label{opt: MOO}
    \min_{\theta} F(\theta) := \left(f_1(\theta), ..., f_M(\theta)\right)
\end{align}
where $f_i(\theta) := \E_{\xi\sim \cD_i}\left[f_i(\theta;\xi)\right]$ is the objective function for the $i$-th task, $M$ is the number of co-trained tasks, $\theta$ denotes the model parameters and $\xi\sim \cD_i$ denotes data sampling process that generates the objective function $f_i$. Moreover, we define the average as $\bar f(\theta) := \frac{1}{M}\sum_{i=1}^{M}f_i(\theta)$. 
In practice, it is usually impossible to obtain a solution $\theta_*$ that minimizes $f_i(\theta)$ simultaneously for every $i$. We thus adopt the notions of optimality and stationarity in MOO literature~\citep{desideri2012multiple, fliege2019complexity}.
\begin{definition}
    A point $\theta\in \R^n$ is called {\bf Pareto optimal}, when there is no $\theta'\in \R^n$ such that $f_m(\theta')\leq f_m(\theta)$ for any $1\leq m \leq M$ and $F(\theta)\neq F(\theta')$. It is called {\bf weak Pareto optimal}, when there is no $\theta'\in \R^n$ such that $f_m(\theta') < f_m(\theta)$ for any $1\leq m\leq M$. A point $\theta\in \R^n$ is called {\bf Pareto stationary}, when $\range(\nabla F(\theta)\T) \bigcap (-\R_{++}^M)$ is an empty set. In other words, there does not exist a vector $v\in \R^n$, such that every coordinate of $\nabla F(\theta)\T v$ is negative. 
\end{definition}
From a theoretical perspective, in general it is impossible to find Pareto optimal points unless we impose some convexity assumptions. Thus it is tempting to consider finding stationary points in training neural networks, which usually operates in the nonconvex regime. To further illustrate how to obtain Pareto stationary points, we consider the following problem, which aims at finding a convex combination of gradients with the minimum norm.
\begin{equation}\label{eq: mgda_d}
    d(\theta)= \sum_{m=1}^{M}\lambda_m^*(\theta)\nabla f_m(\theta),\ \lambda^*(\theta) = \argmin_{\lambda\in \Delta^M}\norm{\nabla F(\theta)\lambda}^2.
\end{equation}
where for any positive integer $s$ we denote by $\Delta^s := \{\lambda\in \R^s| \lambda_i \geq 0,\ \sum_{i=1}^{s}\lambda_i  = 1.\}$ the probability simplex in $\R^s$. We use $\norm{\cdot}$ to represent $\ell^2$ norm for vectors and Frobenius norm for matrices. We use $\norm{\cdot}_2$ to represent spectral norm for matrices. It has been shown that~\citep{fliege2019complexity}, on one hand, if a point $\theta$ is not Pareto stationary, then $\norm{d(\theta)}$ in \eqref{eq: mgda_d} is positive and one can update $\theta$ according to $d(\theta)$ to simultaneously reduce the objective functions. On the other hand, if $\theta$ is Pareto stationary, then $d(\theta) = 0$ in \eqref{eq: mgda_d}. Thus we further define the notion of stationarity in the MOO sense as follows.
\begin{definition}
    We say a point $\theta$ is $\epsilon$-Pareto-stationary in Problem \eqref{opt: MOO}, when $\norm{d(\theta)}\leq \epsilon$ in \eqref{eq: mgda_d}.
\end{definition}

\section{Approach}
\label{sec: Parameter Gradients Balancing is Infeasible}
We aim to achieve Pareto stationary between the co-trained multiple tasks (e.g., prediction of like, share and comment) in industrial multi-task ranking system. The goal is hard to achieve due to gradient conflicts reported by some papers~\citep{sener2018multi,yu2020gradient}. Therefore, we are motivated to design efficient algorithms to balance these task-specific gradients.

Given an industrial-scale MTL ranking system as illustrated in Figure \ref{fig: SparseNN}, how could we balance task-specific gradients to achieve Pareto stationary? 

This question can be divided to two sub-questions:
\begin{itemize}
    \item Key Question 1 (\textbf{KQ1}): How can we efficiently obtain per-task gradients of billions of model parameters (see Table \ref{tab:Proportion of parameters}) under a constrained compute budget?
    \item Key Question 2 (\textbf{KQ2}): Given the extremely high dimensional per-task gradients, how can we balance these gradients to achieve Pareto stationary? 
\end{itemize}

We will present our system design in Section \ref{sec: system_design} as our solution for \textbf{KQ1}. With that in hand, we then propose the gradient balancing algorithmic framework called \texttt{MultiBalance} to address \textbf{KQ2}.

\begin{algorithm}[t]
    \small
	\caption{\texttt{Multi-Task learning parameters update}}
	\label{alg: defult MTL update} 
	\begin{algorithmic}[1]
	    \State \textbf{Input} Initial model parameter $\theta_0$, learning rate $\alpha$.
	    \For {$k=0, \dots, K-1$}
	        \State $f_{sum}(\theta_k) = f_1(\theta_k)+f_2(\theta_k)+\dots+ f_M(\theta_k)$
                \State Call $f_{sum}(\theta_k).backward()$ to obtain $\nabla f_{sum}(\theta_k)$
    	       \State $\theta_{k+1} = \theta_k - \alpha_k \nabla f_{sum}(\theta_k)$
	    \EndFor
	\end{algorithmic} 
\end{algorithm}

\begin{algorithm}[t]
    \small
	\caption{\texttt{M-time backward passes to balance parameter gradients}}
	\label{alg: M-times backward pass} 
	\begin{algorithmic}[1]
	    \State \textbf{Input} Initial model parameter $\theta_0$, learning rates.
	    \For {$k=0, \dots, K-1$}
                \State $V_{k}$ is an empty list
	        \For {$m=1, \dots, M$}
	            \State Call $f_{m}(\theta_k).backward()$ to obtain $\nabla f_{m}(\theta_k)$
                    \State $V_{k}.append(\nabla f_{m}(\theta_k))$
	        \EndFor
                
            \State $V_{k}\prime = Gradient\_ Balancing(V_{k})$
                     
	    \State $\theta_{k+1} = \theta_k - \alpha_k * sum(V_{k}\prime)$
	    \EndFor
	\end{algorithmic} 
\end{algorithm}

\subsection{System Design}\label{sec: system_design}
We first answer KQ1 in this subsection. The default setting of multi-task model training is shown in Algorithm \ref{alg: defult MTL update}. Since all task objective functions are summed together and we call backward function once, we only have $\nabla f_{sum}(\theta)$. However, per-task gradient $\nabla f_m(\theta)$ for $m$-th task are required in gradient balancing. 

Prior work from academia \citep{sener2018multi,yu2020gradient,wang2020gradient,chen2020just,liu2021towards,zhou2022convergence} normally does backward pass $M$ times to get per-task gradients w.r.t. the shared parameters as shown in Algorithm \ref{alg: M-times backward pass}. However, this method suffers from an obvious drawback: time complexity becomes $M$ times as it requires to back-propagate each task loss vs. the default setting only back-propagate the sum of all task losses. In industry, large-scale  system is highly sensitive to time complexity, and training QPS regression by $M$ times is generally unaffordable.

\textbf{Key design.} To overcome the challenge, \texttt{MultiBalance} is to balance the per-task gradients w.r.t. the shared feature representation rather than the shared parameters.

\begin{figure*}
  \centering
 \subfigure[MultiBalance: forward.]{
    \label{fig: MultiBalance-forward}
    \includegraphics[width=0.45\linewidth]{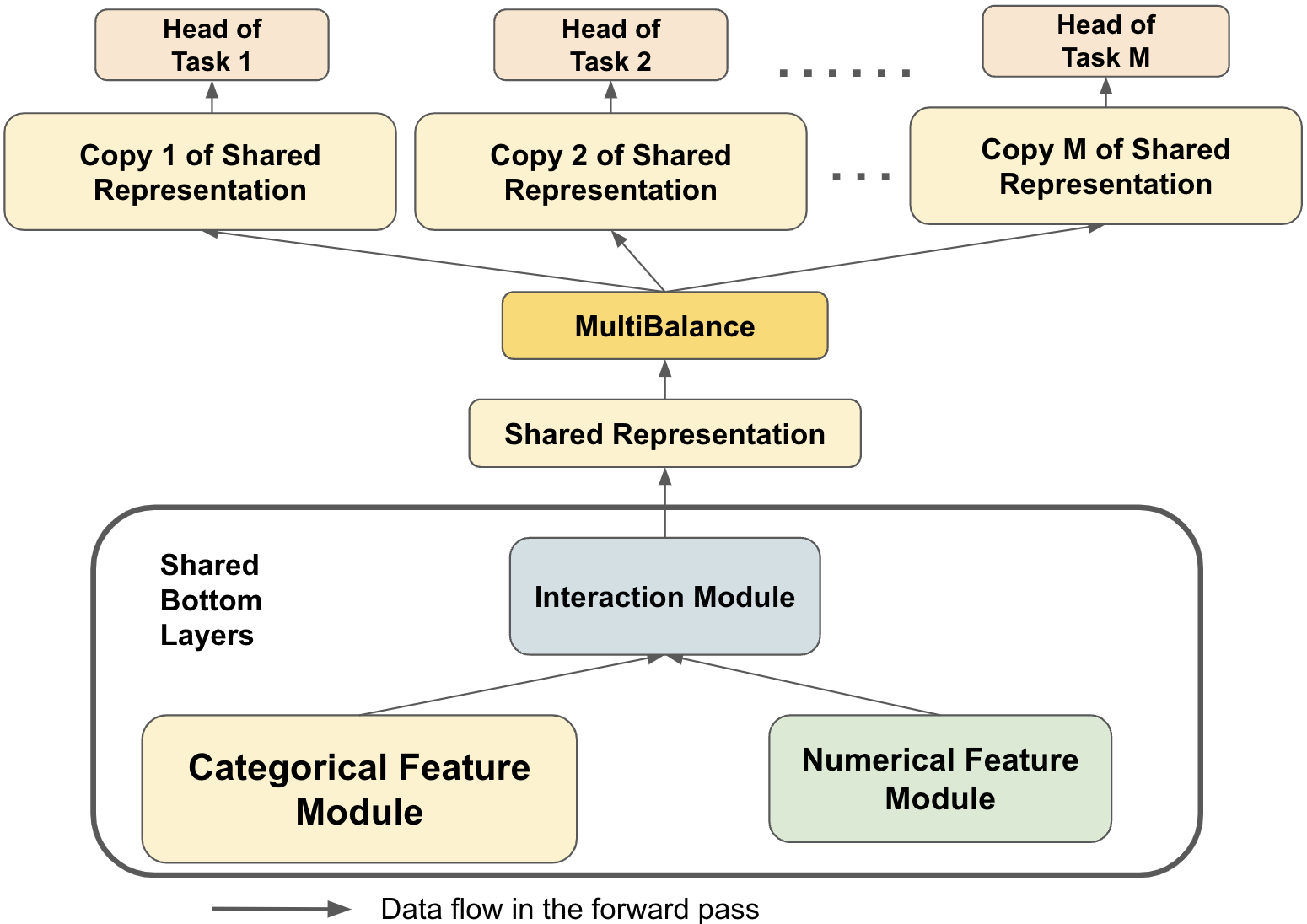}}
 \subfigure[MultiBalance: backward.]{
    \label{fig: MultiBalance-backward}
    \includegraphics[width=0.45\linewidth]{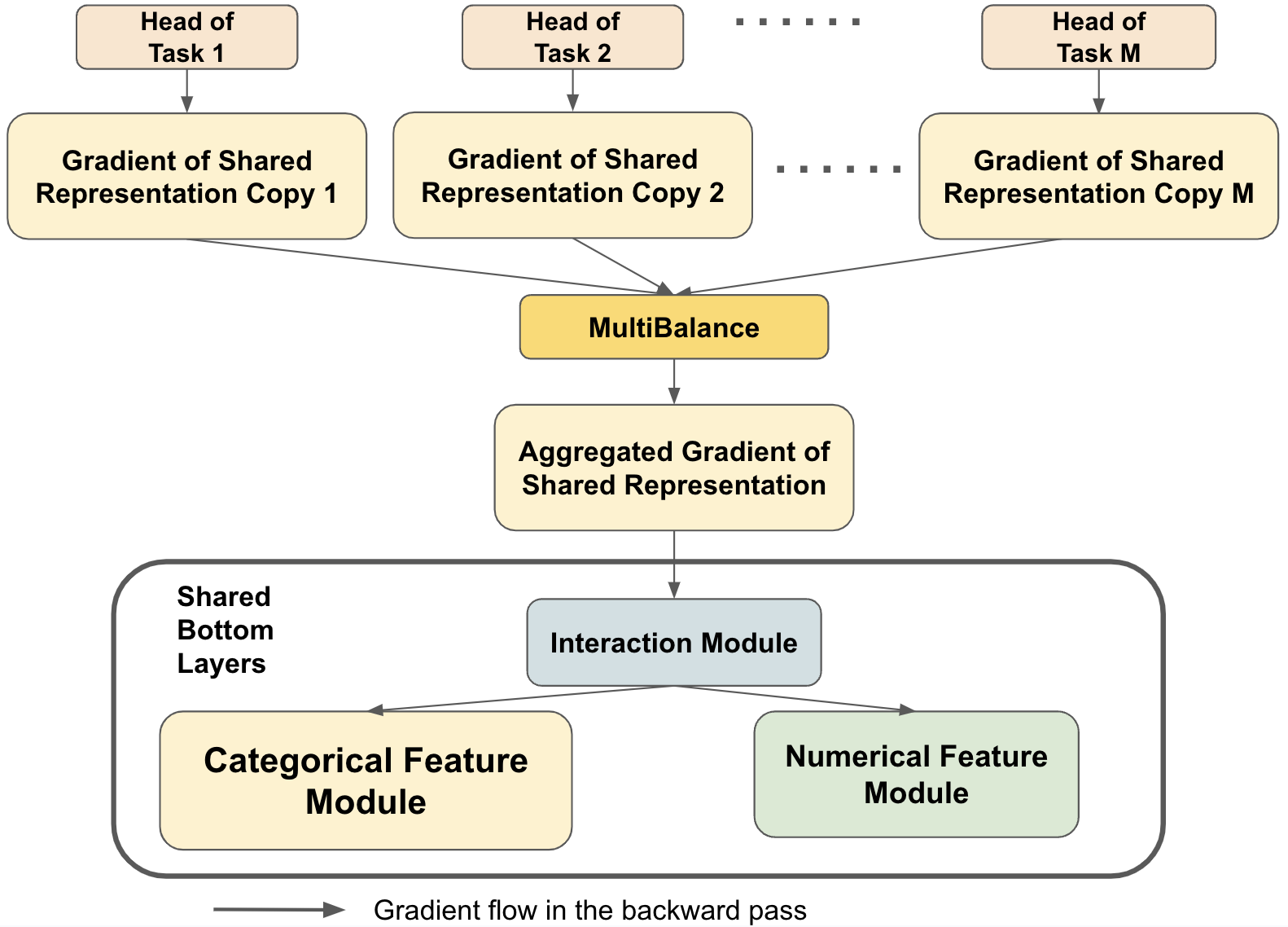}} 
\end{figure*}

\textbf{Forward pass.} \texttt{MultiBalance} copies the shared representation into $M$ copies and each copy is fed into a task head in the forward pass as shown in Figure \ref{fig: MultiBalance-forward}. 

\textbf{Backward pass.} In the backward pass (as shown in Figure \ref{fig: MultiBalance-backward}), the per-task gradients w.r.t. the shared representation (the copy of this task head) will be calculated by automatic differentiation\footnote{\url{https://pytorch.org/tutorials/beginner/basics/autogradqs_tutorial.html}}. \texttt{MultiBalance} will learn a weight vector based on these gradients (refer to as representation gradients in the rest of this paper) and then the learned weights are used to weighted-sum the per-task gradients into an aggregated gradient. After that, the aggregated representation gradient will be returned to autograd engine and continue the current ongoing backward pass. Given the aggregated representation gradient, the gradients of shared parameters will be calculated via automatic differentiation according to the chain rule, which means the balancing effect will be back propagated to all shared parameter. We will elaborate why the learned weights can minimizing all task objectives as much as possible in Section \ref{sec: Direction-based Multi-Objective Balancing}.

{\bf We summarize highlights of MultiBalance as follows:}
\begin{enumerate}
    \item Comparing to existing work that typically requires $M$ backward passes to compute the $M$ task-specific gradients, our proposed method \texttt{MultiBalance} only requires single pass. Crucially, we copy the shared representation to $M$ copies and feed each of them into a task head. Thus, during the only one backward pass, each task-specific gradient is back propagated from the task objective to the corresponding task-specific copy of the representation.
    \item Even after computing $M$ task-specific gradients, balancing representation gradients still has superior performance as compared to that of parameter gradients in terms of computational complexity. Classical \texttt{MGDA}-type methods~\citep{desideri2012multiple, fliege2019complexity, liu2021stochastic, fernando2022mitigating, xiao2023direction, chen2023three} as well as many others like \texttt{PCGrad}~\citep{yu2020gradient} and \texttt{Gradient Vaccine}~\citep{wang2021gradient} require computing $\<\nabla f_i(\theta),\nabla f_j(\theta)>$ for all $(i,j)$ pairs, which leads to $\cO(M^2 n)$ per-iteration complexity, where $n$ denotes the dimension of model parameters. In stark contrast, in Pytorch\footnote{\url{https://pytorch.org/tutorials/beginner/blitz/autograd_tutorial.html}} \texttt{MultiBalance} can be implemented as an intermediate operator of the running autograd graph and hence only requires one backward pass with $\cO(M^2 n')$ per-iteration complexity, where $n'$ is the dimension of the shared representation and is much smaller than $n$ (e.g., $n'$ is usually of order $10^3$ while $n$ is of order $10^9$). The balanced representation gradients will be back-propagated to all shared parameters (i.e., the balancing effect can be back propagated to all shared parameters) via the automatic differentiation (AD).
\end{enumerate}

\subsection{Direction-based Multi-Objective Balancing}
\label{sec: Direction-based Multi-Objective Balancing}
We will answer KQ2 in this section. We first introduce a closely related line of work in MOO literature based on conflict-averse direction~\citep{liu2021conflict}, which shares similarity with \texttt{MGDA}-type methods. Then we explain why it is not directly applicable to handle industrial-scale problems. Finally we present the building blocks of our efficient algorithms with a novel analysis.

\noindent{\bf Preliminary: Conflict-averse direction.} Conflict-averse direction in \cite{liu2021conflict} aims at finding a direction that minimizes all functions as much as possible. In particular, they consider linear approximation of each function (i.e., Taylor expansion up to the linear term) which leads to the following minimum decrease rate function:
\begin{align}
    R(\theta, d) = \min_{1\leq i\leq M} \alpha^{-1}(f_i(\theta) -f_i(\theta-\alpha d)) \approx \min_{1\leq i\leq M}\<\nabla f_i(\theta), d> \label{eq: ca_rate}
\end{align}
where $\alpha$ denotes the learning rate and $d$ is a convex combination of per-task gradients as shown in Equation \eqref{eq: mgda_d}. To make the worst-case decrease as large as possible, we may consider finding a direction that maximizes the above function, while keeping it close to a pre-specified direction obtained from prior knowledge as a regularization. For a pre-specified weight vector $\lambda_0 = (\lambda_{0,1}, ..., \lambda_{0,M})^\top\in \Delta^M$ and $\rho>0$, \texttt{SDMGrad} \citep{xiao2023direction} considers 
\begin{align}
    \max_{d\in \R^n}\min_{1\leq i\leq M}\<\nabla f_i(\theta), d> - \frac{1}{2}\big\|d - \rho \sum_{i=1}^{M}\lambda_{0,i}\nabla f_i(\theta)\big\|^2 \label{opt: smd_obj}
\end{align}
whose solution $d^*$ has the following form
\begin{align}
    d^* = \nabla F(\theta)\lambda_{\rho}^*,\ \lambda_{\rho}^* = \argmin_{\lambda\in \Delta^M}\big\|\nabla F(\theta)\lambda + \rho \sum_{i=1}^{M}\lambda_{0,i}\nabla f_i(\theta)\big\|^2, \label{eq: sdm_weight}
\end{align}
where $\rho\lambda_0$ is the user-defined prior weight vector. Intuitively, problem \eqref{opt: smd_obj} aims at maximizing the worst-case decrease (i.e., \eqref{eq: ca_rate}) while restricting $d$ to not be too far away from the direction $\rho \sum_{i=1}^{M}\lambda_{0,i}\nabla f_i$, by penalizing their $\ell^2$-distance. It is straight-forward from \eqref{eq: sdm_weight} that for $\rho=0$ the algorithmic framework reduces to \texttt{MGDA}. Now to obtain the update direction in \eqref{opt: smd_obj} it suffices to solve for $\lambda_{\rho}^*$ in \eqref{eq: sdm_weight}. Instead of solving for the subproblem \eqref{eq: sdm_weight}, \texttt{SDMGrad} \citep{xiao2023direction} approximates $\lambda_{\rho}^*$ via one-step projected stochastic gradient descent. 

\medskip
\noindent{\bf Representation Gradient.} However, we highlight here again that the dimensionality of $\theta$ is extremely large (see Table \ref{tab:Proportion of parameters}), and thus even one-step projected stochastic gradient as suggested by \citet{xiao2023direction} is unrealistic as it requires $M$ backward passes in a single update. One way to handle this problem is to replace the parameter gradients used in \eqref{eq: sdm_weight} with representation gradients via the system designed in Section \ref{sec: system_design}, on top of this we develop our novel direction-based MOO algorithms based on representation gradients with a novel analysis. To our best knowledge, our theory that validates the surrogate problem \eqref{opt: mgdaub_stochastic}, in particular the decomposition in \eqref{eq: param_grad_w_residual} and the characterization of $R_{\cB}$ are novel.

We now formally define the representation gradient and elaborate how it can be used as a surrogate to learn a weight vector. As shown in Figure \ref{fig: SparseNN}, in multi-task learning, typically the goal is to learn a representation shared by all tasks while maintaining different task-specific heads for different tasks. To explicitly showcase different set of model parameters, we suppose the optimization problem we consider in \eqref{opt: MOO} takes the form
\begin{align}\label{opt: MOO_special}
    \min_{\theta = (W, \phi)} \ell(W,\phi) := \left(\ell_1(W, \phi_1), ..., \ell_M(W, \phi_M)\right)
\end{align}
where we define
\begin{align}
    &F(\Phi(W; x), \phi; y) := (f_1(\Phi(W; x), \phi_1; y), ..., f_M(\Phi(W; x), \phi_M; y)) \notag\\
    &\ell(W, \phi) := \E_{(x, y)\sim \cD}\left[F(\Phi(W; x), \phi; y)\right].
\end{align}
$(x, y)\sim \cD$ denotes the data and label pair sampled from a given distribution $\cD$, which fits both the offline and online training setting. For a particular $(x, y)$, $\Phi(W;x)$ is the representation (parameterized by $W$ and to be learnt) given $x$ as the input. $\phi = (\phi_1, ..., \phi_M)\in \R^{q\times M}$ are the task-specific parameters, i.e., we denote by $\phi_i$ the parameters that only belong to task $i$ for $i=1,...,M$. Moreover, we overload the notation and define
\begin{align*}
    f_m(\Phi(W; x_{\cB}), \phi_m; y_{\cB}) = \frac{1}{|\cB|}\sum_{(x,y)\in \cB}f_m(\Phi(W; x), \phi_m; y) 
\end{align*}
for a given batch of data $\cB=\{(x_i, y_i): i=1, 2,..., |\cB|\}$. Notice that the task-specific parameters $\phi_m$ are owned by $m$-th task only, and thus for the update of $\lambda_k$ in Algorithm \ref{alg: mobile} we can exclude the calculation for the gradients with respect to $\phi$. In other words, for any given batch of data $\cB$ we have $\nabla_{\theta} F\T \nabla_{\theta} F = \nabla_{W} F\T \nabla_{W} F$.
Hence we can focus on analyzing $\nabla_W F$. Note that for a given $\lambda\in \Delta^M$,
\begin{align}
    &\underbrace{\nabla_W F(\Phi(W; x_{\cB}), \phi; y_{\cB})}_{\text{parameter gradients}}\lambda \notag\\
    = &\sum_{m=1}^{M}\lambda_m \nabla_W f_m(\Phi(W; x_{\cB}), \phi_m; y_{\cB}) \notag\\
    = &\sum_{m=1}^{M}\sum_{(x,y)\in \cB} \frac{\lambda_m}{|\cB|}\nabla_W\Phi(W;x)\T\nabla_{\Phi} f_m(\Phi(W; x), \phi_m; y) \notag\\
    = &\E_{(x,y)\sim \cD}\left[\nabla_W\Phi(W;x)\right]\T\underbrace{\nabla_{\Phi} F(\Phi(W; x_{\cB}), \phi; y_{\cB})}_{\text{representation gradients}}\lambda + R_{\cB}, \label{eq: param_grad_w_residual}
\end{align}
where the second equality holds by chain rule, and the residual term $R_{\cB}$ is defined as
\begin{align}
    &r_m(x,y) = (\nabla_W\Phi(W;x) - \E_{(x,y)\sim \cD}\left[\nabla_W\Phi(W;x)\right])\T\nabla_{\Phi} f_m(\Phi(W; x), \phi_m; y), \notag\\
    &R_{\cB} = \sum_{m=1}^{M}\sum_{(x,y)\in \cB} \frac{\lambda_m r_m(x,y)}{|\cB|}. \label{eq: res}
\end{align}
By some concentration inequalities (e.g., Bernstein inequality) we can deduce that for sufficiently large batch size $|\cB|$, with high probability $R_{\cB}$ is negligible. Hence the task weights of one stochastic version of \texttt{MGDA} 
(i,e, \eqref{eq: sdm_weight} with $\rho=0$) can be approximately seen as
\begin{align}\label{opt: mgda_stochastic}
    \min_{\lambda\in\Delta^M} \norm{\E_{(x,y)\sim \cD}\left[\nabla_W\Phi(W;x)\right]\T\nabla_{\Phi} F(\Phi(W; x_{\cB}), \phi; y_{\cB})\lambda}.
\end{align}
Note that the representation gradients $\nabla_{\Phi} F$ are relatively cheaper to compute as compared to the parameter gradients $\nabla_W F$, since the former only requires a partial backward pass while the latter requires a whole pass. Therefore in practice~\citep{sener2018multi}, one may solve the following problem as a surrogate.
\begin{align}\label{opt: mgdaub_stochastic}
    \min_{\lambda\in\Delta^M} \norm{\nabla_{\Phi} F(\Phi(W; x_{\cB}), \phi; y_{\cB})\lambda} .
\end{align}
We further combine the idea in \texttt{SDMGrad} to introduce a prior task weight $\lambda_0$ to update $\lambda_k$ in each iteration -- in line \ref{eq: lambda} of Algorithm \ref{alg: mobile} we perform one-step projected gradient descent, i.e., 
\begin{align}\label{eq: lambda_update}
    \lambda_{k+1} = \Pi_{\Delta^M}(\lambda_{k} - \beta_{k}V_{k+1}\T (V_{k+1}\lambda_{k} + \rho V_{k+1}\lambda_0)),
\end{align}
where $V_{k+1}$ collects the representation gradients of all tasks, $\Pi_{\Delta^M}$ denotes the projection onto the probability simplex $\Delta^M$ and $\beta_k$ denotes the learning rate. After updating the task weight, we set the average (weighted by $\lambda_{k+1}$) of the shared representation gradients from different tasks as an aggregated gradient of the shared representation, and then pass it to the bottom level of the modules to update model parameters via optimizers like SGD~\citep{robbins1951stochastic} or Adam~\citep{kingma2014adam} (lines \eqref{eq: repregrad_replace} and \eqref{eq: theta_opt} of Algorithm \ref{alg: mobile}), as also depicted in Figure \ref{fig: MultiBalance-backward}.

\begin{algorithm}[t]
    \caption{\texttt{MultiBalance}}
    \label{alg: mobile} 
    \begin{algorithmic}[1]
        \State \textbf{Input} Initial model parameter $\theta_0$, learning rates, prior task weights.
        \For {$k=0, \dots, K-1$}
                \State $V_{k+1}$ is an empty list. \label{eq: vk_init}
                \State Apply one backward pass over a batch $\cB$ to obtain 
                \[
                    g_{k+1, m} =\nabla_{\Phi} f_m(\Phi(W_k; x_{\cB}), \phi_k; y_{\cB}), 1\leq m\leq M.
                \]
            \For {$m=1, \dots, M$} 
                \State $u_{k+1,m} = (1 - \gamma_{k, m})u_{k, m} + \gamma_{k, m}\norm{g_{k+1, m}}$.
                \State $v_{k+1,m} = \frac{u_{k+1, m}}{\norm{g_{k+1, m}}} g_{k+1, m}$. 
                \State $V_{k+1}.append(v_{k+1, m})$.
            \EndFor
         \State $\lambda_{k+1} = \Pi_{\Delta^M}(\lambda_{k} - \beta_{k}V_{k+1}\T (V_{k+1}\lambda_{k} + \rho V_{k+1}\lambda_0))$. \label{eq: lambda}
         \State Replace representation gradient with $V_{k+1}\lambda_{k+1}$ and continue the current ongoing backward pass. \label{eq: repregrad_replace}
         \State Update $\theta_k$ via SGD or Adam to get $\theta_{k+1}$.\label{eq: theta_opt}
        \EndFor
        \State \textbf{Output} $\theta_K$
    \end{algorithmic} 
\end{algorithm}

\medskip
\noindent{\bf Stabilizing Stochastic Gradients.} Empirically, we observe that the magnitudes of stochastic gradients fluctuate, and thus directly using them to perform \eqref{eq: lambda_update} as suggested by \citet{xiao2023direction} leads to unstable training. To stabilize the magnitudes of stochastic gradients (denoted as $g_{k, m}$), we follow \citet{malkiel2020mtadam} to keep track of the moving average (denoted as $u_{k, m}$ for $m$-th task) of the norms of the gradients, and then re-scale the gradients based on such an average as follows.
\begin{align*}
    u_{k+1,m} = (1 - \gamma_{k, m})u_{k, m} + \gamma_{k, m}\norm{g_{k+1, m}},\ v_{k,m} = \frac{u_{k, m}}{\norm{g_{k, m}}} g_{k, m}.
\end{align*}
With this we complete the design of Algorithm \ref{alg: mobile}. The seemingly complex updates in Algorithm \ref{alg: mobile} can be divided into two parts -- model training (line \eqref{eq: theta_opt}) and task balancing (lines \eqref{eq: vk_init} - \eqref{eq: repregrad_replace}). From this perspective, the implementation of our algorithm is simple and straightforward as it only introduces an extra balancing step as compared with vanilla optimization algorithm.

\medskip
\noindent{\bf Theoretical Results}
Now we quantitatively characterize the difference between optimization via parameter gradients in \eqref{opt: mgda_stochastic} and via representation gradients in \eqref{opt: mgdaub_stochastic}.
\begin{lma}\label{lem: mgda_ub}
    Suppose we are given matrices $A\in \R^{q\times n}, B\in \R^{p\times q}$ and constants $\ell \geq \mu\geq 0$ satisfying $\mu^2 I \preceq B^\top B\preceq \ell^2 I$.
    Define $\lambda_{X, *} = \argmin_{\lambda\in \Delta^n}\norm{X\lambda}$
    for any matrix $X\in \R^{d\times n}$. Then we have
    \begin{align*}
        \mu^2 \norm{A\lambda_{A, *}}^2 \leq \mu^2\norm{A\lambda_{BA, *}}^2\leq  \norm{BA\lambda_{BA, *}}^2\leq \norm{BA\lambda_{A, *}}^2 \leq \ell^2 \norm{A\lambda_{A, *}}^2.
    \end{align*}
\end{lma}
The proof is straightforward, by noticing that the first and the third inequalities use the definitions of $\lambda_{BA, *}$ and $\lambda_{A, *}$, and the second and the fourth inequalities use $\mu^2 I \preceq B^\top B\preceq \ell^2 I$. Note that in the extreme case when $\mu = \ell$, we have $B^\top B=\mu^2 I= \ell^2 I$ and all inequalities in Lemma \ref{lem: mgda_ub} become equalities. This indicates that solving for $\min_{\lambda\in\Delta^n}\norm{BA\lambda}$ is nearly the same as solving for $\min_{\lambda\in\Delta^n}\norm{A\lambda}$ when the spectrum of $B^\top B$ spans a short range (i.e., when $\mu$ is close to $\ell$). This Lemma also indicates that solving the problem $\min_{\lambda\in\Delta^n}\norm{A\lambda}$ can be seen as optimizing an upper bound for $\min_{\lambda\in\Delta^n}\norm{BA\lambda}$. This directly leads to the following result.

\begin{theorem}\label{thm: upperbound}
    Suppose there exist $\lambda\in \Delta^M$ and $\ell, \epsilon, \delta>0$ such that
    \begin{align*}
        &\big\|\E_{(x,y)\sim \cD}\left[\nabla_W\Phi(W;x)\right]\big\|_2\leq \ell,\ \norm{\nabla_{\Phi} F(\Phi(W; x_{\cB}), \phi; y_{\cB})\lambda}\leq \epsilon,\ \norm{R_{\mathcal{B}}}\leq \delta,
    \end{align*}
    where $R_{\cB}$ is in \eqref{eq: res}. Then 
    $\norm{\nabla_W F(\Phi(W; x_{\cB}), \phi; y_{\cB})\lambda} \leq \ell \epsilon + \delta$.
\end{theorem}
Note that for sufficiently large batch size $|\cB|$ the norm of the residual term $\norm{R_{\mathcal{B}}}$ can be well controlled (say bounded by $\delta$). Then according to Theorem \ref{thm: upperbound}, we know that as long as we find a weight vector $\lambda$ such that the objective in \eqref{opt: mgdaub_stochastic} is bounded by $\epsilon$, then $(W,\phi)$ is a $(\ell\epsilon + \delta)$-Pareto-stationary point on batch $\cB$. 

\section{Experiments}
In this section, we evaluate \texttt{MultiBalance} on Meta's industrial-scale ads ranking model and two feeds ranking models over real world training and test examples.


\subsection{Experimental Goals and Baselines}
\textbf{Efficiency Goal.} The first goal is to evaluate the training efficiency of balancing representation gradients and parameter gradients in industrial-scale systems. Since \texttt{MultiBalance} is a MOO method, to be fair, we compare it with another two MOO methods that balance per-task gradients w.r.t. shared parameters. They are: (1) \texttt{MGDA}~\citep{sener2018multi}, which learns a weight vector in the probability simplex $\Delta^M$ and use this vector for re-weighting task gradients $\nabla f_m(\theta)$ for $m=1,...,M$ to get the aggregated gradient $d(\theta)$ for updating $\theta$; and (2) \texttt{MoCo}~\citep{fernando2022mitigating}, a state-of-the-art MOO method, which outperforms existing \texttt{MGDA}-type methods because it maintains a moving-average of gradients to calculate the weight vector, and thus helps mitigate the bias in $d(\theta)$ caused by variance in stochastic gradients. We will show that \texttt{MultiBalance} is much more efficient than the methods that balance parameter gradients in subsection \ref{sec: Representation Gradient vs. Parameter Gradient}


\noindent\textbf{Performance Goal.} Given that balancing representation gradients is the way to pursue, the second goal is to compare \texttt{MultiBalance} with these public baseline methods: (1) \texttt{Uncertainty}~\citep{kendall2018multi}, which maximizes log-likelihood of the model prediction uncertainty; (2) \texttt{PCGrad}~\citep{yu2020gradient}, which replaces one per-task gradient by its projection onto the normal plane of another task gradient if the two gradients are conflicting (i.e., $\<\nabla f_i(\theta), \nabla f_j(\theta)> < 0$); (3) \texttt{Gradient Vaccine}~\citep{wang2020gradient}, which encourages two per-task gradients to have a similar angle with their historic angle; (4) \texttt{Gradient Drop}~\citep{chen2020just}, which drops a per-task gradient with a probability when this gradient has the opposite sign with the other per-task gradients, which is to avoid tug-of-wars between gradients; (5) \texttt{DB-MTL}~\citep{lin2023scale}, which encourages all task gradients to have a similar gradient magnitude (maximum, minimum, median or mean of all task gradients.) as a target gradient. We evaluated these choices and find out median is the most stable one and report the corresponding result; (6) \texttt{IMTLG}~\citep{liu2021towards}, which optimizes the task weights via a closed-form solution, such that the aggregated gradient (sum of weighted per-task gradients) has equal projections onto individual gradient. \textit{To be fair, we also apply previous non-MOO methods to balance representation gradients} and we observe that \texttt{MultiBalance} can outperform all baseline methods in Section \ref{sec: Comparison with Baseline Methods}.


\subsection{Experimental Setup}
We note that existing study on MOO-based methods for MTL is mostly on public dataset~\citep{sener2018multi, liu2021conflict}, and clearly there is a huge discrepancy between the industrial datasets and the publicly available ones, which makes them unsuitable for downstream experiments. In this paper, we conduct experiments on Meta's production model with an industrial dataset that consists of billions of samples. We design the experiments to understand how our proposed algorithms can improve multi-task learning and we do not touch user-specific attributes for privacy compliance.

\textbf{Training Efficiency Metric.} Model inference only incurs forward passes without gradients for balancing, and we mainly focus on improving the training efficiency. QPS denotes the number of training samples consumed by the model per second. Hence, the QPS difference between the model with or without gradient balancing methods indicates the time cost of the methods. 


\textbf{Performance Metric.} Normalized entropy (NE) is defined as:
\begin{equation}
NE = \frac{\frac{-1}{n}\sum_{i=1}^{n}(y_{i}\log(p_i)+(1-y_{i})\log(1-p_{i}))}{-(p\log(p)+(1-p)\log(1-p))}.
\label{equ:NE}
\end{equation}
Here $n$ represents the total number of examples in the dataset. $y_i \in {0, 1}$ are the labels, $i = 1, 2,\dots n $ and $p_i$ is the estimated probability of a user behavior (e.g., click an ad) for each impression, while $p$ is the average empirical probability of the user behavior. It measures how accurately a model is predicting when users will click on ads. Lower is better. Note that NE is a single metrics that is evaluated against all training/evaluation examples instead of any specific user or user group. NE difference is obtained by: $NE(vanilla \ MTL) - NE(MTL \ with \ gradient \ balancing)$ and thus the positive number means NE gain and negative number means NE loss.

\textbf{Tasks.} In ads ranking domain, it is well known \citep{ma2018entire, wen2020entire} that post-view click (CTR), post-click conversion (CVR) and post-view conversion (Conv|imp) are the tasks and we co-train them on one model as shown in Figure \ref{fig: SparseNN}; Due to privacy policy, tasks in feeds models are not released but readers could refer to publications like \citet{wu2022feedrec} that tasks are normally the predictions of user behaviors on a feed such as ``like'', ``share'' and ``comment'' etc. We report the average NE diff of all feeds tasks. There are two feeds models A and B as we have different surfaces (e.g., Facebook or Instagram) at Meta.

\begin{table*}[htbp]
  \centering
  \setlength{\abovecaptionskip}{0.0cm}
  \setlength{\belowcaptionskip}{0.0cm}
  \small
  \caption{Comparison between Balancing Shared Parameter Gradients and Shared Representation Gradients}
    \begin{tabular}{cccccc}
    \toprule
    \multirow{2}[0]{*}{Methods} & \multicolumn{1}{c}{\multirow{2}[0]{*}{Target to be balanced}$^{\dagger}$}
    & Training Efficiency & CVR   & Conv|imp & CTR \\
          & \multicolumn{1}{c}{} & QPS Drop (\%) & Eval NE Gain & Eval NE Gain & Eval NE Gain \\
    \midrule
    \multirow{3}[0]{*}{\texttt{MGDA}} & Interaction Module & 82.353 & 0.062 & -0.010 & -0.021 \\
          & Numerical Feature Module & 68.235 & -0.015 & 0.000 & -0.007 \\
          & Categorical Feature Module & N/A$^{*}$ & N/A   & N/A   & N/A \\
    \midrule
    \multirow{3}[0]{*}{\texttt{MoCo}} & Interaction Module & 82.524 & 0.033 & -0.028 & -0.028 \\
          & Numerical Feature Module & 73.370 & -0.035 & -0.013 & -0.016 \\
          & Categorical Feature Module & N/A$^{*}$ & N/A   & N/A   & N/A \\
    \midrule
    \textbf{\texttt{MultiBalance}} & Shared Representation & \textbf{0.42} & 0.091 & -0.014 & -0.032 \\
    \bottomrule
    \end{tabular}%
    \\
    {\raggedright  $*$ Training QPS of parameter gradients balancing on categorical feature module drops too much to get a feasible training job. Note that categorical feature module accounts over 90\% parameters of a model (see Table \ref{tab:Proportion of parameters}). \par}
    {\raggedright $\dagger$ The gradients of the corresponding target to be balanced. For example, ``Shared Representation'' in the last row of this column indicates the gradients of the shared representation will be balanced in \texttt{MultiBalance}. \par
    } 
  \label{tab:MoCo on ads model}
\end{table*}

\subsection{Balancing Representation Gradients vs. Balancing Parameter Gradients}
\label{sec: Representation Gradient vs. Parameter Gradient}
In this subsection, we demonstrate that balancing the gradients of shared representations like \texttt{MultiBalance} has much higher ROI (Return On Investment) than balancing the gradients of shared parameters. `Return' here indicates the performance improvement while `investment' means the extra cost of gradient balancing. We present the numerical results in Table \ref{tab:MoCo on ads model}. Note that it has been shown that \texttt{SDMGrad} can be seen as \texttt{MGDA} with regularization (see, e.g., Section 4.1 of \cite{xiao2023direction} for details), we thus only include \texttt{MGDA} as a representative algorithm of this type of methods.


\textbf{Balancing Parameter Gradients has low ROI.} As shown in Table \ref{tab:MoCo on ads model}, if we balance per-task gradients w.r.t. the interaction module or numerical feature module, the cost will be around 70\%-80\% for \texttt{MGDA} and \texttt{MoCo}, which is huge as we consider QPS decrease more than 5\% as a significant loss. This huge QPS drop is understandable: compared with vanilla MTL~\citep{caruana1997multitask} (i.e., all task weights are equal and add up to $1$) which only requires one backward pass, they require $M$ backward passes to obtain per-task gradients w.r.t. the shared parameters. Similarly, it is not surprising that parameter gradients balancing on categorical feature module is too slow to obtain a feasible training flow given that the categorical module accounts for 90\% of the parameters as discussed in Section \ref{sec: MTL_industry}. The investment (70\%-80\% QPS drop) is huge; nevertheless, the return is limited: \texttt{MGDA} has 0.062\% gain while \texttt{MoCo} has 0.033\% gain on CVR. Given that interaction module only has a very small part of model parameters in industrial system (Table \ref{tab:Proportion of parameters}), the limited performance is understandable. If we can balance the parameters within categorical feature model, a better gain is possible. However, balancing gradients of the parameters within categorical feature model is too time-consuming to have a feasible flow. 

\textbf{\texttt{MultiBalance} has high ROI.} In comparison, the cost of our algorithm \texttt{MultiBalance} is nearly neutral with merely 0.42\% QPS drop, which is much more efficient than \texttt{MGDA} and \texttt{MoCo}. Meanwhile, we observe that \texttt{MultiBalance} improves CVR largely (0.091\% NE gain, and the gains over 0.05\% are considered `significant') with neutral impact on Conv|imp and slight loss (0.032\% NE loss) on post-view CTR. E-commerce companies like Amazon\footnote{https://getida.com/resources/blog/advertising/the-golden-metrics-that-drive-success-on-amazon/} treat CVR more important than CTR as CVR measure the percentage of the ads that convert to real purchase. To conclude, we see that in stark contrast to parameter gradients balancing that require time-consuming computation, our proposed method \texttt{MultiBalance} is easier to implement, achieves more efficient balancing, and has relatively higher ROI.

\begin{table*}[htbp]
  \centering
    \setlength{\abovecaptionskip}{0.0cm}
  \setlength{\belowcaptionskip}{0.0cm}
  \small
  \caption{Experimental Results. Metric is NE Gain on the Eval Set}
    \begin{tabular}{c|c|c|cccc}
    \toprule
    Domain & Feeds A & Feeds B & \multicolumn{4}{c}{Ads} \\
    \midrule
    Tasks & Ave of All Tasks & Ave of All Tasks & CVR   & Conv|imp & CTR   & Ave of All Tasks \\
    \midrule
    \texttt{Uncertainty} & 0.291 & 0.488 & 0.111 & -0.154 & -0.200 & -0.081 \\
    \texttt{PCGrad} & 0.200 & 0.213 & -0.100 & -0.154 & -0.222 & -0.159 \\
    \texttt{Gradient Vaccine} & NEX$^{*}$   & 0.175 & -0.016 & -0.026 & 0.000 & -0.014 \\
    \texttt{Gradient Drop} & -0.092 & 0.362 & -0.079 & -0.090 & -0.062 & -0.077 \\
    \texttt{DB-MTL} & 0.390 & 0.522 & 0.048 & -0.103 & -0.112 & -0.056 \\
    \texttt{IMTLG} & -0.273 & 0.110 & 0.063 & -0.090 & -0.162 & -0.063 \\
    \texttt{MultiBalance} remove EMA & 0.476 & -0.122 & -0.007 & -0.124 & -0.143 & -0.091 \\
    \texttt{\textbf{MultiBalance}} & \textbf{0.476} & \textbf{0.738} & 0.091 & -0.014 & -0.032 & \textbf{0.015} \\
    \bottomrule
    \end{tabular}%
    \\
    {\raggedright  \enspace $*$ Loss explosion consistently occurs when we apply Gradient Vaccine at this model. \par}
  \label{tab:Experimental Results}%
\end{table*}%


\subsection{Comparison with Baseline Methods}
\label{sec: Comparison with Baseline Methods}
The experimental results are presented in Table \ref{tab:Experimental Results}. First of all, we  calculate the average NE diff of all tasks at ads and feeds models and observe that \texttt{MultiBalance} achieves the best overall performance across the three models and is the only one with average NE gains at ads model. The detailed results of ads model has been discussed in the last subsection. We discuss on the experimental results of feeds ranking model here, where we observe that \texttt{MultiBalance} achieves the best performance of 0.476\% NE gain at feeds model A and the best performance of 0.738\% NE gain at feeds model B, largely outperforming all baseline methods. A side interesting observation is that all gradient balancing methods achieve much better results on Feeds than Ads. A possible reason is that the tasks in Feeds have more common knowledge to enhance each other than Ads and we leave this to future study.

Among baseline methods, in the ads model, we observe that \texttt{Uncertainty}, \texttt{DB-MTL} and \texttt{IMTLG} can also bring gains on CVR but they also hurt the performance of the other two tasks largely. Therefore, the solution obtained by \texttt{MultiBalance} is better than all baseline methods in terms of Pareto efficiency. In the feeds models, we observe that \texttt{Uncertainty}, \texttt{DB-MTL} and \texttt{PCGrad} consistently improve the performance across the two models and show superiority than other prior work.

\begin{figure*}
  \centering
\setlength{\abovecaptionskip}{0.0cm}
  \setlength{\belowcaptionskip}{-0.2cm}
 \subfigure[Gradient Magnitudes before Balancing]{
    \label{subfig: Gradient Norm before Balancing}
    \includegraphics[width=2.0in]{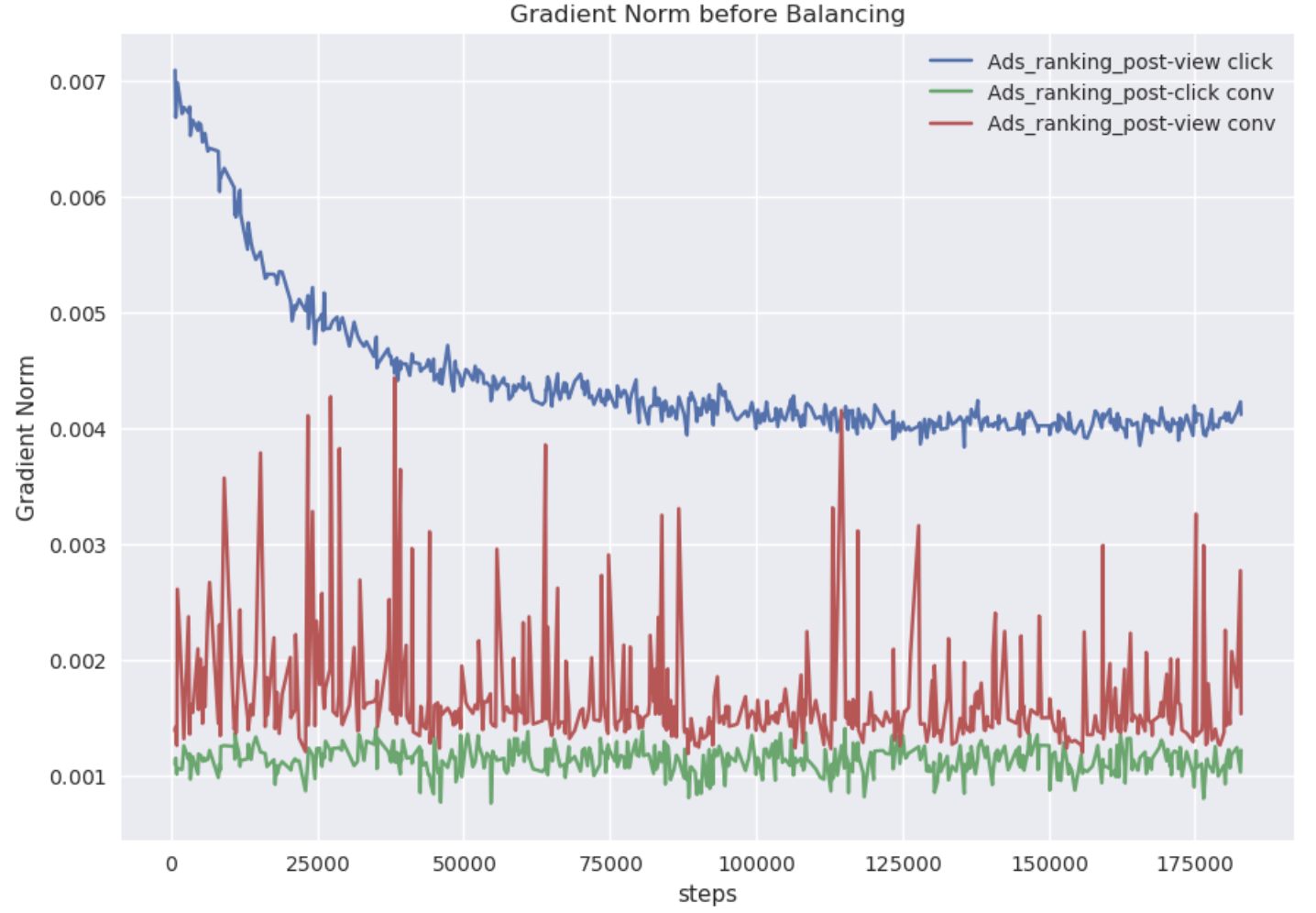}}
 \subfigure[Task Weights Learned by MultiBalance]{
    \label{subfig: Task Weights Learned by MultiBalance} 
    \includegraphics[width=2.0in]{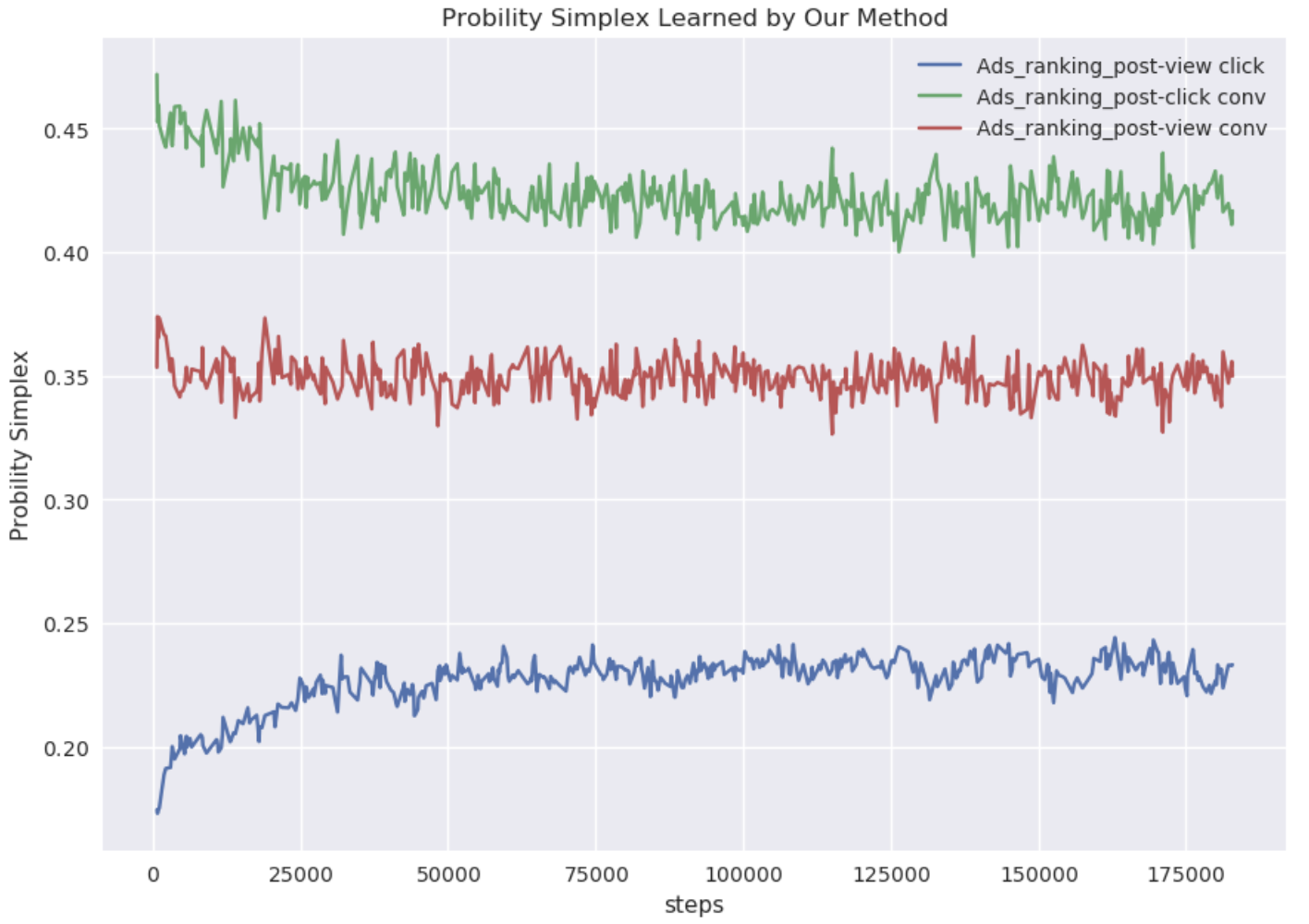}} 
 \subfigure[Gradient Magnitudes after Balancing]{
    \label{subfig: Gradient Norm after Balancing} 
    \includegraphics[width=2.0in]{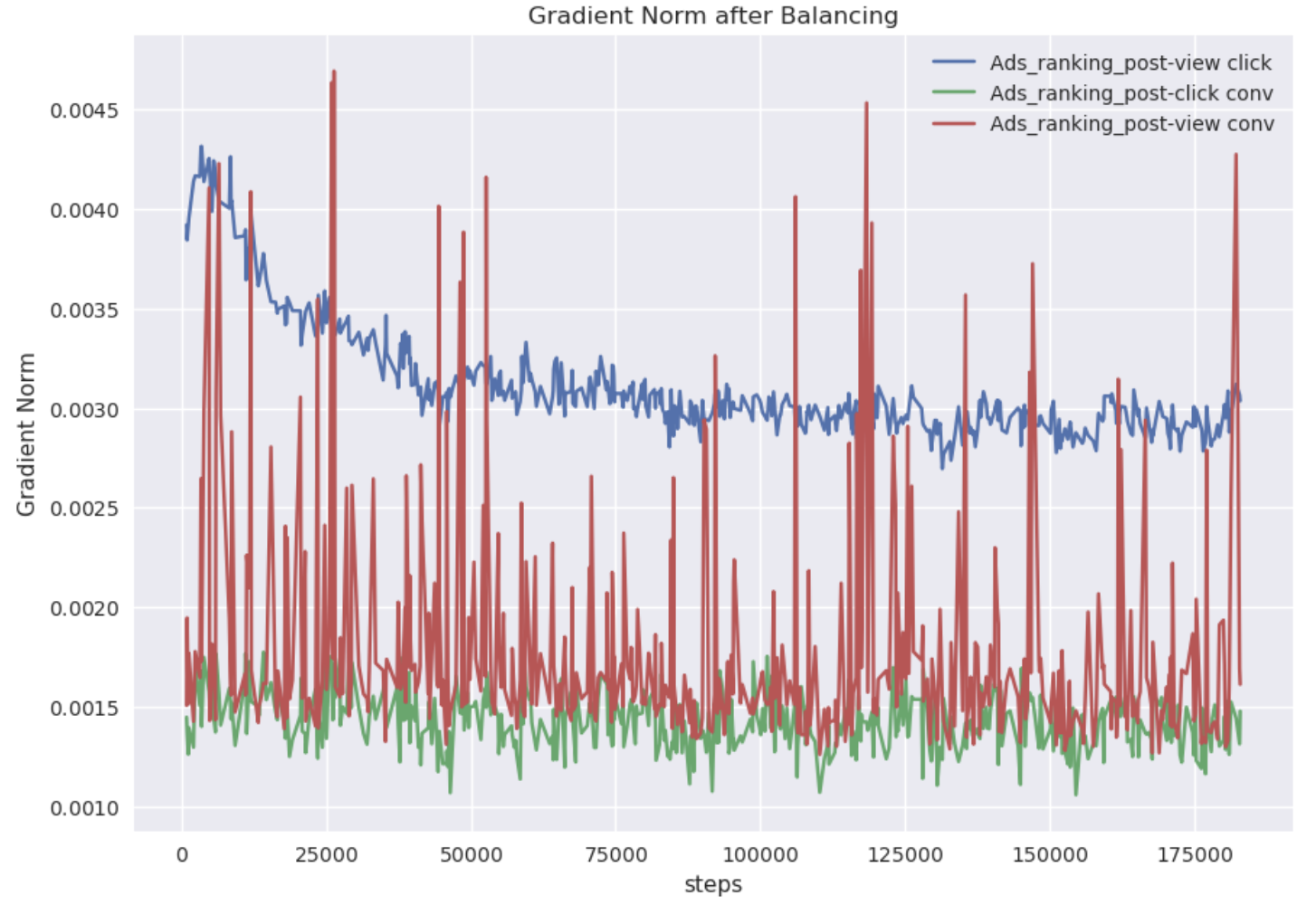}}   
   \caption{The Visualization of Learning Process of MultiBalance. Green is CVR, Blue is CTR and Red is Conv|imp.}%
  \label{fig: visualization of multibalance} 
\end{figure*}

\textbf{Visualization.} We also present the visualization of the learning process of \texttt{MultiBalance} in Figure \ref{fig: visualization of multibalance}. First, in Figure \ref{subfig: Gradient Norm before Balancing}, we observe that CTR's original gradient magnitude (blue) is much larger than CVR's (green) and Conv|imp's (red) and hence dominate the optimization. Figure \ref{subfig: Task Weights Learned by MultiBalance} shows the task weights learned via \texttt{MultiBalance} where the solution converges to around 0.25, 0.4 and 0.35 for CTR, CVR, Conv|imp. In other words, lower weight is assigned to CTR to alleviate its dominance over the other two tasks, which can be observed in Figure \ref{subfig: Gradient Norm after Balancing} that the task gradients multiplied with the corresponding weights has a more balanced magnitude to each other.

\textbf{The impact of moving average of gradient magnitudes.} To stabilize the magnitudes of
stochastic gradients, we keep track
of the moving average of the gradient magnitudes and re-scale the gradients based on such an average. We note that this technique introduces minor additional memory cost as the moving-average term is only a one-dimensional scalar. In Table \ref{tab:Experimental Results}, we compare the performance between \texttt{MultiBalance} with and without such component. We observe that this moving average can significantly improve the performance on the ads model and feeds model B. We also observe that baseline methods without such moving average encounter instability issue like loss explosion (NEX) of \texttt{Gradient Vaccine} in Feeds model A.

\textbf{The impact of learning rate.} The impact of learning rate ($\beta$ in Algorithm \ref{alg: mobile}) to learn the task weights is shown in Table \ref{tab:The impact of learning rate feeds}. In practice we notice that the update direction for $\lambda_k$, i.e., $V_{k+1}\T (V_{k+1}\lambda_{k} + \rho V_{k+1}\lambda_0)$, has relatively small magnitudes, and thus we replace it with the cosine similarity between $V_{k+1}$ and $V_{k+1}\lambda_{k} + \rho V_{k+1}\lambda_0$. On top of that, we {also observe that the learning rate has a huge impact on the performance. In the ads model, the larger learning rate is, the better performance of CVR is achieved and the more Conv|imp and CTR are hurt. From the perspective of Pareto stationary, we select 1.0 as the best learning rate as significant improvement is achieved for CVR while prevent significant losses for other tasks. In the Feeds model, we observe that an appropriate learning rate is also important to achieve Pareto stationary and it shows that learning rate around 5.0 is the best in this specific model.

\textbf{The impact of other hyperparameters.} We do not heavily tune other hyperparameters of our methods. For prior task weights $\lambda_0$ and its coefficient $\rho$ in Equation \eqref{eq: lambda_update}, we simply set $\lambda_0 = \frac{1}{M}(1, ..., 1)\T$ and $\rho=0.1$.

\begin{table}[htbp]
  \centering
  \setlength{\abovecaptionskip}{0.0cm}
  \setlength{\belowcaptionskip}{0.0cm}
  \small
  \caption{The impact of learning rate on feeds models}
    \begin{tabular}{ccc}
    \toprule
    Tasks & Feeds model A & Feeds model B  \\
    \midrule
    Metrics (\%) & Average Eval NE Gain & Average Eval NE Gain \\
    \midrule
    learning rate = 1.0  & 0.34 & 0.38\\
    learning rate = 2.0  & 0.30 & 0.59\\
    learning rate = 5.0  & 0.48 & 0.74\\
    learning rate = 10.0  & 0.41 & 0.55 \\
    \bottomrule
    \end{tabular}%
  \label{tab:The impact of learning rate feeds}%
\end{table}%
\section{Conclusions}
Multi-task learning is very important for industrial recommendation systems. However, multi-task learning often suffer from negative transfer. In this paper, we propose a novel and practical approach called \texttt{MultiBalance} to balance per-task gradients with respect to the feature representation shared by all tasks. Experimental results shows large improvements with neutral training efficiency cost. We also analyze the theoretical characteristics of feature representation gradients and show it is a good surrogate of parameter gradients. We also give a condition when \texttt{MultiBalance} is provably to achieve Pareto stationary point.

\newpage

\bibliographystyle{abbrvnat}
\bibliography{paper.bib}

\appendix

\end{document}